\documentclass[%
 aip,
 amsmath,amssymb,
 preprint,%
]{revtex4-1}

\usepackage{graphicx}
\usepackage{dcolumn}
\usepackage{bm}

\usepackage[utf8]{inputenc}
\usepackage[T1]{fontenc}
\usepackage{mathptmx}

\begin{document}

\preprint{AIP/123-QED}

\title{Towards low gas consumption of muographic tracking detectors in field applications}

\author{G. Nyitrai}
\email{nyitrai.gabor@wigner.hu}
\affiliation{ Wigner Research Centre for Physics, 1525 Budapest, Pf. 49., Hungary }
\affiliation{ Budapest University of Technology and Economics, 1521 Budapest, Pf. 91., Hungary }

\author{G. Hamar}

\author{D. Varga}
\homepage{http://regard.wigner.hu}
\affiliation{ Wigner Research Centre for Physics, 1525 Budapest, Pf. 49., Hungary }

\date{\today}

\begin{abstract}

Gaseous detectors are widely used in high energy physics, and are attractive choices in tracking systems for cosmic muon imaging, also called muography. Such detectors offer high resolution and high efficiency at reasonable cost for large sizes, however, one of the drawbacks is that the gaseous detection medium must be prevented from contamination by outside air or internal outgassing. Standard systems work with a constant gas flow, leading to regular maintenance in the form of gas cylinder changes, which can be an issue for remote field applications. In this paper we discuss the practical possibilities to reduce gas consumption of an outdoor gaseous tracker, where particularly the gas density change from daily temperature cycling limits the input flow. Such "breathing" effect can be circumvented by well designed buffer volume, which must prevent external air contamination. A realistic MWPC tracking test system with 0.9 square meter area, total volume of 160 l, has been operated for 36 days with a flow of 3 l/day, confirming that the buffer volume, in this case a 50 m long and 10 l volume low diffusion tube, ensures sufficient gas quality. The key effects governing the gas flow dynamics, including diffusion and gas volume change, has been studied quantitatively, leading to practical design prescriptions.

\end{abstract}

\maketitle

\section{\label{sec1}Introduction}

Imaging the internal structure of large size objects with cosmic particles, or muography in short, is a non-invasive survey method, initiated in the 1960-ies, with pioneering work by Alvarez \cite{ALVAREZ} searching for hidden chambers in a pyramid, and later applied successfully on magma dynamics of volcanoes by Tanaka \cite{TANAKA_2014}. It is used worldwide, including major volcanoes such as Etna \cite{DOMENICO_2020},  Vesuvius \cite{MURAVES_2014, MURAVES_2018}, Stromboli \cite{STROMBOLI_2019}; La Soufrière volcano \cite{MARTEAU_2016, MARTEAU_2017, MARTEAU_2019}; and Sakurajima \cite{SMO_2018}. Muography has developed so that the imaging properties and possible methodologies are now well understood. A large number of  applications exist \cite{MURAVES_2020_PPNP, TANAKA_2018} such as search for archaeological or natural underground structures \cite{ECHIA_2017, ECHIA_2019, SURDA_2012, THOMPSON_2020, TANAKA_2020}, mining  \cite{SCHOUTEN_2018}, nuclear reactor imaging  \cite{REACTOR_2013} and renewed interest in the internal structure of large historical objects \cite{BIGVOID_NATURE_2017}.

Muography requires a tracking system for precision directional measurement of muons, for which three basic technologies exist. Emulsions \cite{BIGVOID_EMULSION, ECC_2015} offer low maintenance and simple installation, at the cost of pre- and post-processing of the detection material, and lack the possibility of real time imaging. Scintillators \cite{MURAVES_2020_JINST, DOMENICO_2018} are probably the most popular systems being highly reliable and needing low maintenance, however these may have high weight, and become expensive if high angular resolution is required.

The third class of detector technologies apply gaseous detection medium. Such detectors offer high efficiency, good position resolution, low weight, and are cost efficient \cite{MWPC_2016}. However, besides their complexity, the usage of gas as detection medium presents a challenge. The outdoor application in muography, requiring low maintenance, tolerance of environmental effects and robustness, gave rise to recent developments in instrumentation which can reduce gas consumption and ensures safe and reliable operation.

The family of "gaseous tracking detectors" is very broad, including the classical Multi-Wire Proportional Chambers (MWPCs)\cite{SAULI} but also recent developments for improved position resolution and rate capability, called the Micro-Pattern Gaseous Detectors \cite{BIGVOID_MPGD, LAZARO_2020}, and Resistive Plate Chambers (RPC) \cite{PUYDEDOME_MURAY+RPC, TOMUVOL} particularly for improved time resolution. The technical challenges are also very diverse, and there are design choices with corresponding pros and cons.

 For any gaseous detector, sufficient gas purity must be maintained, since the working gas can be contaminated by air (oxygen) by either diffusion or outgassing from construction materials  \cite{BLUMROLANDI} which reduces performance. For all discussions below, we define "outgassing" as gain reduction relative to the pure gas (high flow) case. Throughout the paper, the "gain" is defined as the mean of the signal amplitude for identified tracks. This is always expressed relative to a reference: normalized either to the gain of other detectors, or the gain relative to a reference time period.

In fact, outgassing is a very broad and complicated term, which includes electronegative molecules from construction materials, but also ambient oxygen from diffusion through detector walls, or water which changes electron drift velocity.  An inspiring study \cite{Why_do_we_flush} pinpointed specific construction materials as source of water and oxygen contamination in muon detectors, and demonstrated the effectiveness of a partially recycled, filtered gas system. Contamination causes various adverse effects such as ageing \cite{BLUMROLANDI}, which are in our cases limited due to the low natural radiation level.

The "outgassing", as here defined, includes not only diffusion, but even tiny leaks if exist. The contamination can be translated to effective air content, which would correspond to the same gain reduction \cite{MWPC_2016}. The outgassing rate can be quantified from the time derivative of the gain reduction, assuming closed system with constant environmental conditions ("sealed mode").

Indeed it would be advantageous to make a gaseous detector fully sealed, and various studies made progress in this direction. In fact it is possible to construct a large size cosmic detector using sealed proportional counters \cite{INDIA_UHE_2005}. A portable muon telescope was proposed using gas-tight RPC-s \cite{WUYCKENS_2018}, whereas under laboratory conditions as much as half a year sealed operation was demonstrated \cite{LOPES_2020}.

On the practical side, construction is easier and weight can be lower if some gas flow is allowed to purge contamination. This is so because daily pressure variation and volume variation can be large, therefore the system is preferably open-ended to avoid mechanical stress on the containment structure. Certainly the supply flow should be as small as possible.

The key question which we address in this paper: how to achieve as low input flow as possible in a gaseous detector for a realistic, full size portable outdoor muography system?

Challenges related to a low-flow system in outdoor conditions include the following:

\begin{itemize}
    \item Intrinsic detector outgassing. The present paper is not dealing with the problem, however, the detector should be as good as possible already in laboratory conditions. The test system presented in this paper was checked for outgassing (gain drop in sealed mode), as discussed below.
    
    \item Day/night temperature variation causes gas volume change, hence "breathing", which can contaminate the gas line from the end (an effect which is much less relevant for underground experiments).
    
    \item Ambient pressure change causes similar "breathing", which may be episodic such as during a storm.
    
    \item Back-diffusion of ambient air from such an open end may cause contamination, therefore a sufficiently long exhaust tube is needed. Placing a barrier such as a bubbler is not applicable, since the "breathing" effect moves typically much larger volume than the supply from the input.
    
\end{itemize}

The problems are most apparent for large gas volumes, which are again, difficult to reduce without performance loss (reducing volume at given area means reducing detector chamber thickness, leading to construction issues \cite{BLUMROLANDI}). As an example, at the Sakurajima Muography Observatory, one single module \cite{VCI_2020} has typically 100-160 l volume, and with a total daily input of 30-40 l, volume change time constant is 3-5 days. 

Adding a buffer volume, which has a sufficiently large volume, and at the same time long enough to restrict diffusion, is a very practical solution. This possibility will be investigated throughout this paper in details. 

The paper is organized as follows. Section \ref{sec2} is about chamber outgassings and tubing radial diffusions. These are the limiting factors running at any gas flow. Section \ref{sec3} elaborates on the diffusion along the buffer tube from the open end to quantify the required length that restrict this effect. Section \ref{sec4} is quantification of flow rate caused by "breathing" from ambient temperature cycling. Section \ref{sec5} discusses the interplay of all these effects, based on a long term outdoor measurement, and demonstrating low flow operation. Section \ref{sec6} presents the practical implementation at the Sakurajima Muography Observatory, and an example calculation of underground application.

\section{\label{sec2}Outgassing and radial diffusion}

Gas contamination, including diffusion through chamber walls and gas release from internal components, is a fundamental limiting factor for detector operation. In our case, this will be measured by the gain reduction according to the definition of the "outgassing" above, since due to the short drift length it correlates well with the tracking performance.

The diffusion along a tube will be called "axial diffusion" (longitudinal diffusion), which actually transfers external air from open end, and can as well transfer contamination between interconnected detector chambers.

Low flow operation requires a gas system with sufficient quality, so that the input gas contamination does not compromise performance. In our case, as it will be shown, this was an unexpectedly large, but acceptable effect due to a long connecting tube between gas source and the first detector chamber input.

\subsection{\label{sec2.2}Measurement of detector outgassing}

\begin{figure}
\includegraphics[width=8cm]{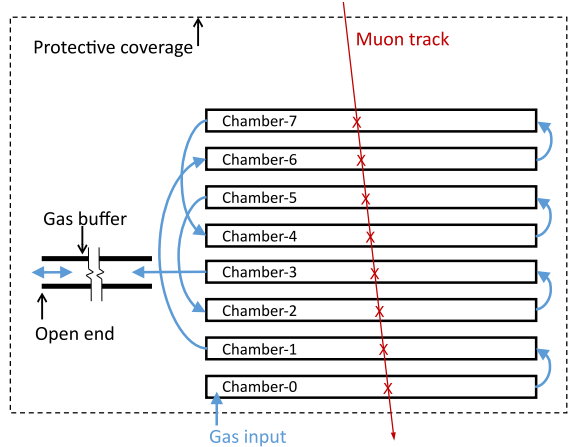}
\caption{\label{setup_cartoon} Scheme showing the detector setup with 8 MWPC chambers. The gas flows serially through the chambers, followed by a buffer tube with open end.
}
\end{figure}

The general measurement setup used throughout the paper is shown in Fig. \ref{setup_cartoon}, with horizontal MWPCs\cite{MWPC_2016}, and a single gas line in series, with input at the bottom layer, using the same readout system as presented in Varga\cite{VCI_2020}. 

The first series of measurements were aimed at quantifying the outgassing of the chambers: that is, the rate of gain reduction in sealed mode. For this, eight chambers with 120 cm $\times$ 80 cm size were installed under laboratory conditions. For sufficiently long time, constant gas flow was driven through all the chambers. After reaching stable gain, the gas line was rearranged so that some of the detector chambers along the chain were isolated (input closed) whereas others still received the constant flow. This way, the latters could be used as references, relative to which the gains of the plugged detector chambers were measured.

\begin{figure*}
\includegraphics[width=17cm]{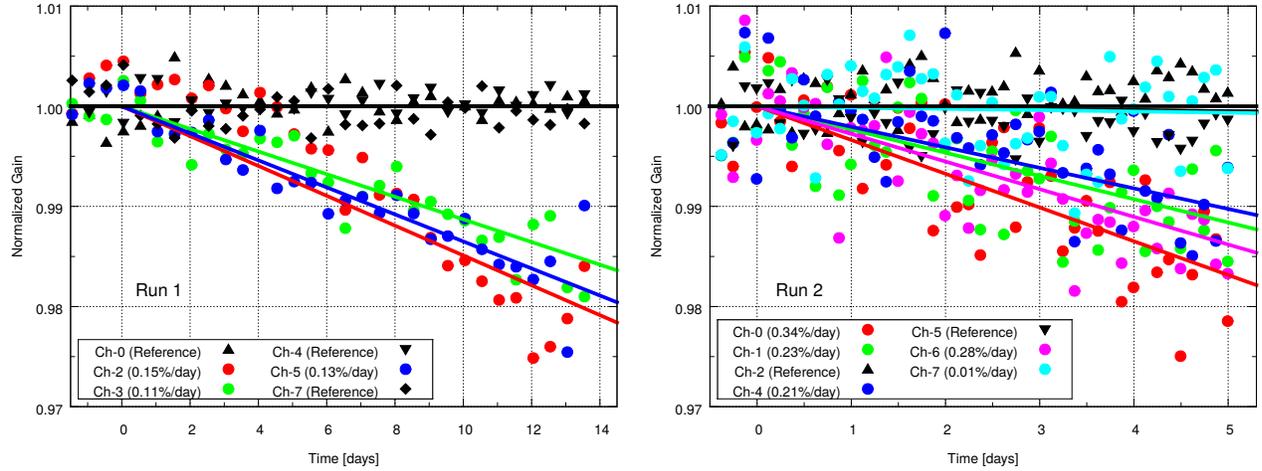}
\caption{\label{fig_Outgassing} Measurement of detector outgassing (gain drop in sealed mode) under laboratory conditions. Out of the 8 detector chambers, some are used as “references” with constant gas flow, relative to the mean of those the gain drop of the others (at zero flow) was measured for 5-12 days. Left and right panels correspond to Run 1 and 2, respectively.}
\end{figure*}

The result of the measurements is shown in Fig. \ref{fig_Outgassing}. In the first Run (left panel) the chambers 2, 3, and 5 were plugged and measured for outgassing with three reference chambers, and in the second Run (right panel) the chambers 0, 1, 4, 6, and 7 were measured with two references. In the course of the two measurements, all detector chambers were characterized, with results summarized in Table \ref{table1}.

\begin{table}[h!]
\centering
\begin{tabular}{|c c c|} 
 \hline
 Chamber & Gain drop (\%/day) & Run \\
 \hline
 Ch-0 & 0.34 & (2) \\ 
 Ch-1 & 0.23 & (2) \\ 
 Ch-2 & 0.15 & (1) \\ 
 Ch-3 & 0.11 & (1) \\ 
 Ch-4 & 0.21 & (2) \\ 
 Ch-5 & 0.13 & (1) \\ 
 Ch-6 & 0.28 & (2) \\ 
 Ch-7 & 0.01 & (2) \\ 
 \hline
\end{tabular}
\caption{Gain reduction rate due to outgassing ( \%/day) for all detector chambers.}
\label{table1}
\end{table}

The conclusion of the measurement is that the outgassing of these units under laboratory conditions is surprisingly low, below 0.3\% per day. Even 30\% gain reduction causes no efficiency loss which means that the full volume can be exchanged in 100 days if we assume linearity. In our case (8 detector chambers with total volume of 160 l) this means a practical lower limit of flow of 0.07 l/h, or 1 cc/min, or 1.6 l/day.

Motivated by this attractively low outgassing of the single chambers, the question we ask is what needs to be done in order to approach this theoretical gas consumption in practical, ambient outdoor conditions? As discussed in the Introduction, if we wish to use a buffer volume which feeds gas back to the chamber system, candidates for such buffer tubing need to be characterized for outgassing as well, which is discussed in the next section.

\subsection{\label{sec2.3}Tubing radial diffusion}

Tubes in the gas system contribute to contamination because of the diffusion through their walls, hence the name "radial". Potential practical choices need to be tested and selected for acceptable outgassing. We have found that some samples of PVC tubes are strongly reducing gain; however, PU (polyurethane), PA (polyamide), PE (polyethylene), copper, and stainless steel perform well. In order to quantify "outgassing", the following measurements were performed.

\begin{figure}
\includegraphics[width=8cm]{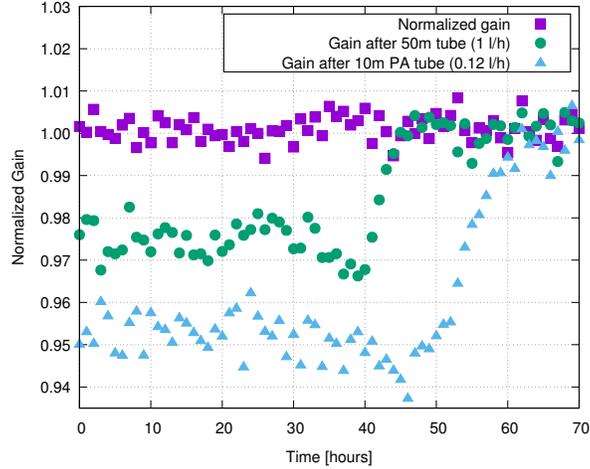}
\caption{\label{fig_RadialDiff} Measurement of the saturated gain drop after a 50 meter long PE-RT tube with 16 mm inner diameter, running at gas flow of 1 l/h, as well as a 10 m long 4 mm inner diameter PA tube at 0.12 l/h flow. One of the reference chambers is indicated in purple. After 40 and 50 hours respectively, the test tubes were removed and the gains clearly returned to the reference.
}
\end{figure}

Using the standard setup with a serial gas line running at constant flow (with a same MWPC detector type but smaller, 40 cm $\times$ 40 cm size chambers), the tube section under study was connected between two chambers. In this setting the chambers before the test tube act as references, and the gain reduction can be measured using the detector chamber after the test tube. In order to further reduce the systematic errors, after reaching the saturated constant gain, the test tube was removed and the original (minimum length) gas connection was re-installed, resulting in a fast return to the reference gain value. Examples are shown in Fig. \ref{fig_RadialDiff}, both for the stability of the reference, as well as the level of gain drop after the test tubes. The lengths and the flows were chosen as they typically occurred during further measurements.

In conclusion, specific gas tube samples could be selected for sufficiently low outgassing, including a small (4 mm inner, 6 mm outer diameter) PA, and a large (16 mm inner, 21 mm outer diameter) PE-RT tube, thus ensuring that gas contamination is low even if the gas dwells for a long time in these tubes.

\section{\label{sec3}Axial diffusion}

The process of air diffusion in one dimension along tubing, most relevantly from the open end, is referred to below as axial diffusion. The process is well understood theoretically, therefore there are very clear predictions about what to expect. A set of measurements were designed in order to test if the system quantitatively follows these predictions, to ensure that in a complicated ambient condition one clearly understands the governing effects and to demonstrate that there are no additional effects beyond the pure diffusion.


The measurement setup consisted of six detector chambers of the same type as in the previous section, note again, with 40 cm size to reduce measurement time and effects related to volume change. The principle is the following: a test tube of variable length was attached to the exhaust of the last chamber in the gas chain, and at $t=0$ the gas input of this single chamber was closed. Due to diffusion, the gas in the exhaust tube starts to be contaminated with air, which at some point reaches the detector. Then air diffuses into the sensitive volume, leading to decrease in the gain. The variable length test tube was rather thick (16 mm inner diameter), and it was attached to a short exhaust tube of fixed 14 cm length and 4 mm inner diameter.

\begin{figure}
\includegraphics[width=8cm]{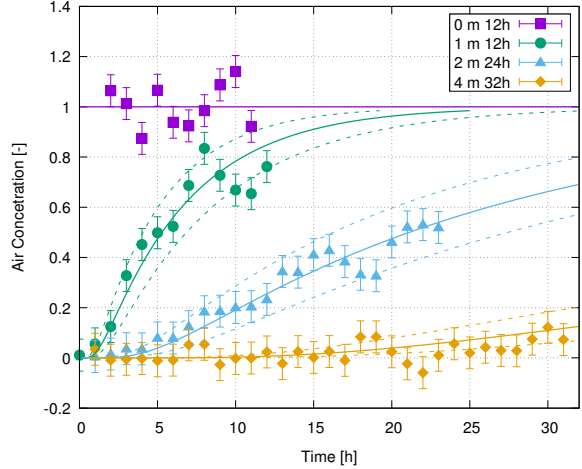}
\caption{\label{fig_AxialDiff4} Air concentration at the detector end of the variable length exhaust tubes, obtained from the gain change rate (normalized time derivative) of the last detector chamber. 0 m length corresponds to a negligible length of the connecting tube, therefore the gain change rate is constant (normalized to 1). The exhaust tubes were connected to the same setup, and the gas supply was closed at $t=0$. Continuous lines show prediction with a $D\;\pm$ 30\% bracket.}
\end{figure}

In the case where the test tube length is practically zero, there is a constant diffusion from the outside ambient air through the short exhaust tube, and therefore one expects a constant gain change (constant time derivative of the gain). This constant corresponds to the "pure air" concentration. The measurement, shown in Fig. \ref{fig_AxialDiff4}, confirms this expectation: gain change rate at the "0 m" test tube length is indeed constant (normalized to 1 on the figure). If one varies the tube length, then the gain change (using the same normalization) effectively measures the air concentration at the end of the tube. Indeed, Fig. \ref{fig_AxialDiff4} shows that 1 m, 2 m, and 4 m test tube lengths feature a considerably slower gain change. The rate of change approaches the ambient concentration after a long time.

The laws of diffusion (Fick's laws) give a very clear analytic prediction for this measurement configuration. In the general case air concentration ($c$) along the tube is governed by the drift-diffusion equation

\begin{equation}\label{eq_driftdiffusion}
    \frac{\partial c}{\partial t} = D \cdot \frac{\partial^2 c}{\partial x^2} - \frac{\partial(c\cdot v)}{\partial x} + D_R,
\end{equation}

where $D$ is the diffusion coefficient, $v$ is the flow velocity in the tube, and $D_R$ is the radial diffusion. The radial diffusion has been discussed in the previous section, and found to be negligible for this measurement. The gas movement along the tube will be the subject of the next section, however for this measurement setup one can assume $v=0$ now. After these considerations, the solution of the pure diffusion equation for air concentration ($c$) in an infinitely long tube is 

\begin{equation}\label{eq_diffusion}
    c = 1 - \mathrm{erf} \left( \frac{L_{diff}}{2\sqrt{D \cdot t}} \right),
\end{equation}

where L\textsubscript{diff} is the length from the open end. For our measurement, the boundary condition at the detector end (opposite to the open end) of the test tube is such that the flux is nearly zero (the short exhaust tube is much thinner than the test tube). Using the relevant solution, shown with continuous lines in Fig. \ref{fig_AxialDiff4}, one gets a nice agreement with the measurement at the diffusion constant value $D =$ 20.1 $\pm$ 0.7 mm$^2$/s (error is statistical only, whereas systematic errors up to 30\% could not be ruled out). It is remarkable to observe the scaling behaviour: a 4 m long tube takes 16 times longer time to reach the same concentration by diffusion. There is no indication of deviation from the "diffusion-only" behaviour after 30 hours, due to outgassing, radial diffusion, or leaks. In practice, due to diffusion only, after 24 (6) hours the air contamination in a tube, originally filled with pure gas, is around 1250 ppm at a distance of 6 (3) m from the open end, calculated from Eq. \ref{eq_diffusion}.

\section{\label{sec4}"Breathing" effect from temperature cycling}

Assuming that the pressure and temperature inside the chambers follow the external conditions, the total mass of working gas stored varies strongly, due to both temperature changes and external pressure changes. In field applications the daily temperature cycles are a major issue at low input flow: if the input does not feed sufficient working gas, external air may be drawn in through the exhaust. In this section, we present the quantification and confirmation of the appearance of this effect.

\begin{figure}
\includegraphics[width=8cm]{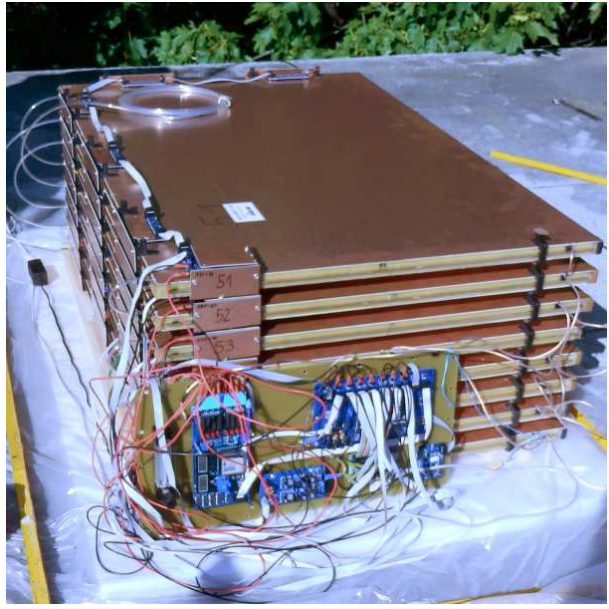}
\caption{\label{fig_roof_setup} The measurement setup installed outdoor, with light protective coverage removed. The structure is the same as the one used for outgassing tests.}
\end{figure}

The image of the measurement setup, following the structure according to Fig. \ref{setup_cartoon}, is shown in Fig. \ref{fig_roof_setup} installed on the flat rooftop of the laboratory building. A light box, as well as reflective coverage, was used to protect from rain and direct sunlight. The eight detector chambers of 80 cm x 120 cm size, each with 20 liter volume, were operated from a variable flow gas source, fed from the inside of the building. As presented in Section \ref{sec2.2}, the ordering of the chambers in the serial gas line was such that the gas entered at the bottom, then fed the top chambers, with the middle chambers at the end of the chain (exhaust after Chamber 3), this way compensating for possible temperature gradient effects inside the box. 

\begin{figure}
\includegraphics[width=9cm]{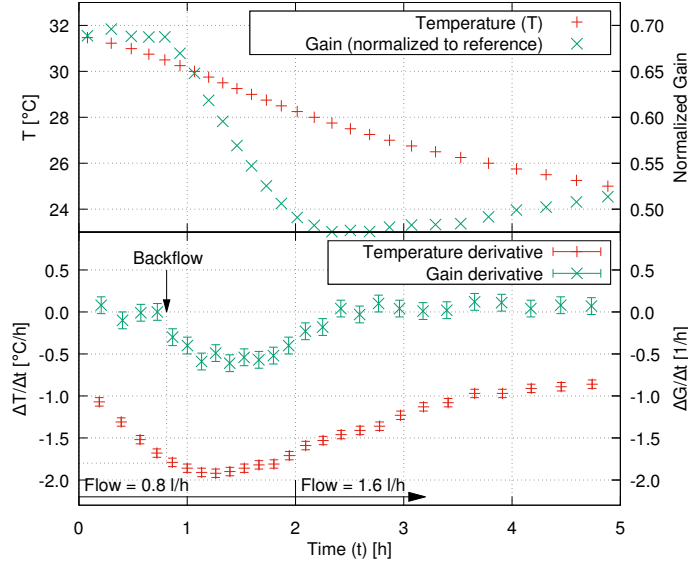}
\caption{\label{fig_Backflow_Time} Measurement of gain change on the last chamber of the gas chain during a period of gradual cooling of the system, at 0.8 l/h flow. The top panel indicates the relative gain and temperature values, the bottom panel shows the time derivative of both. The arrow indicates the onset of the air ingestion, in this case at temperature reduction rate of -1.8 °C/h. After observing the gain drop, the flow was increased to 1.6 l/h to avoid extensive contamination.
}
\end{figure}

When the temperature drops, basic thermodynamics laws predict at which point the exhaust flow $I_{out}$ (expressed in units of l/h) becomes negative, at given input flow $I_{in}$ and temperature change:

\begin{equation}\label{thermo}
    I_{out} = I_{in} + V_{0} \cdot  \frac{ \Delta T' - \Delta P'}{ \Delta t},
\end{equation}

where $T'=(T-T_0)/T_0$ and $P'=(P-P_0)/P_0$ are the relative (Kelvin) temperature and pressure values (assuming that the relative pressure change is small), against fixed $T_0$ and $P_0$ references, and $V_0$ is the total gas volume in the detector at reference state. Thus the equation means that the exhaust flow turns negative if the volume change - due to relative temperature or pressure change over time - is greater than the input flow. In order to test this prediction, we have measured the gain of the last chamber in the chain with a short, negligible volume buffer tube attached to it. We expect a sharp gain reduction when the $I_{out}$ turns negative, and thus air rapidly enters into the last chamber. The measurement is shown in Fig. \ref{fig_Backflow_Time}, confirming a fast gain change at a specific value of the time derivative of the temperature. For clarity, the top panel shows the raw measurement (gain and temperature), whereas the lower panel is the numerical time derivative, which carries the quantitative information (according to Equation \ref{thermo}).

\begin{figure}
\includegraphics[width=8cm]{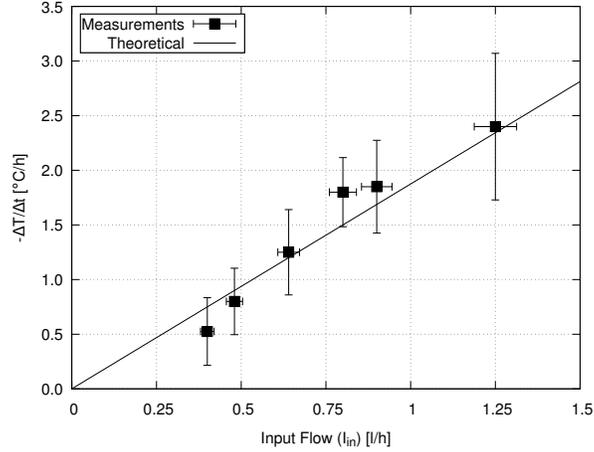}
\caption{\label{fig_Phase} The critical temperature reduction rates at different flow values, where a sharp gain drop was observed due to air ingestion ($I_{out}<0$). The expected slope is determined by Eq. \ref{thermo}, well matching the observation.}
\end{figure}

The measurement was repeated for multiple days, attempting to observe the gain drop, and determine the corresponding critical temperature drop rate for various flow values. Fig. \ref{fig_Phase} shows the measurements with well identified gain breaking points as a function of flow, which indeed follow the direct prediction, and more importantly, confirm the quantitative parameters of our system.

One can conclude that temperature cycling (specifically, temperature reduction) can cause air contamination from the exhaust side which can and must be mitigated by sufficient buffer volume. Combined with the results of the previous section, the buffer tube must be extended with an additional length to compensate for axial diffusion.

\section{\label{sec5}Low-flow outdoor operation}

The test system was the same as the one described in the previous section, and a sufficiently long buffer tube has been attached after the last chamber in the gas line. The required buffer volume can be derived from Eq. (\ref{thermo}) if we take an upper estimation with zero input flow:

\begin{equation}\label{eq_buffer}
    V_{buffer} = V_{0} \cdot \left(T'_{max} - T'_{min} \right).
\end{equation}

A 50 m long PE-RT tube with 16 mm inner diameter has been chosen, that is, nearly 10 liters total volume. A 15 °C temperature drop would correspond to a 8 l volume to be supplied from the buffer tube (40 m) at very low input flow, with sufficient remaining length (10 m) to compensate for daily axial diffusion. The purpose of the test was to demonstrate the practical applicability of the previous considerations.

\begin{figure}
\includegraphics[width=9cm]{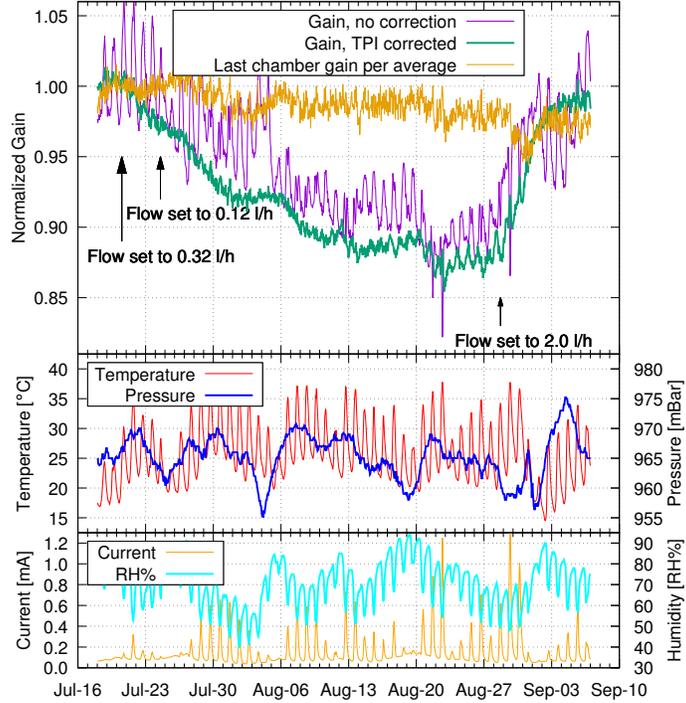}
\caption{\label{fig_Lowflow_AllCH+TP+IH} Top panel: average normalized gain of all chambers (relative to the gain during the first 3 days), over 50 days of measurement. Middle and bottom panel: environmental parameters and anode current. The flow settings are indicated by arrows, the flow was constant at 0.12 l/h during 36 days.
}
\end{figure}

The measurement took about 50 days, during which the system was running continuously. The working gas was supplied from the inside of the building for simpler and better controlled gas input. The overview of the measurement results and conditions is shown in Fig. \ref{fig_Lowflow_AllCH+TP+IH}. The gain is defined as the mean of the signal amplitude for identified tracks, and the "normalized gain" is the gain relative to the mean of the first 3 days. The figure shows the environmental conditions on the middle panel, with considerable daily temperature swings, as well as rainy periods with high humidity.

The flow was 2 l/h during the first 3 days, which was constant for a sufficient time before the beginning of the measurements. Then, for 4 days, the flow was set to 0.32 l/h. As the gain did not show signs of deterioration, the flow was optimistically reduced as low as 0.12 l/h. For 36 days, the system was running at this input flow, when it was raised back to 2 l/h in order to confirm return to the original gain value.

\begin{figure*}
\includegraphics[width=16cm]{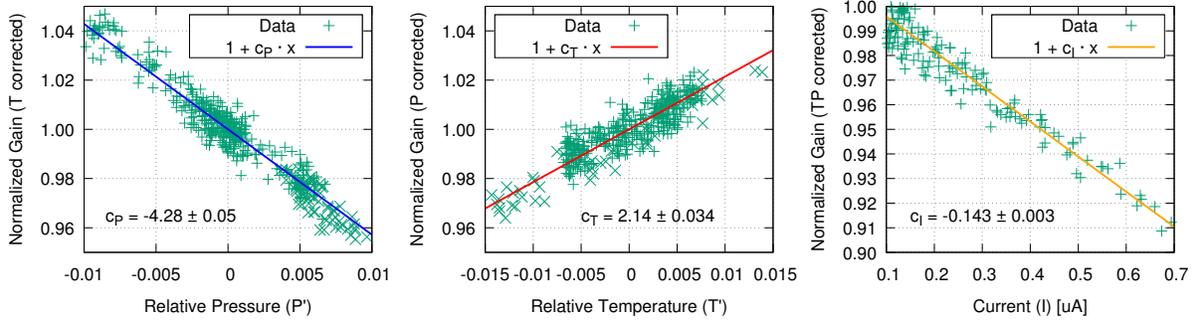}
\caption{\label{fig_TPIcorr} Relative pressure (P'), relative temperature (T') and current (I) correlation of the measured gain, with the straight line fits used for the correction. 
}
\end{figure*}

The daily gain variation due to the correlation with pressure and temperature is evident from the raw data (as the uncorrected gain data shows in Fig. \ref{fig_Lowflow_AllCH+TP+IH}, purple line), caused by gas density change\cite{BLUMROLANDI, SAULI}. In order to correct for this, the correction factors have been determined, as shown in Fig. \ref{fig_TPIcorr}. On the left panel, the normalized measured gain change depends linearly on the the relative pressure change, where $P' = (P - 966)/966$ (here $P$ is measured in mbar). The middle panel shows the correlation with the relative temperature change, where $T' = (T - 293)/293$ (here $T$ is measured in Kelvin). We have identified an additional variable which affects the measured gain: this is the current ($I$) drawn on the high voltage (anode) as it decreases the operating voltage due to the high output resistance of the used HV unit. This dark current is originated from temperature-dependent resistance of the construction materials of these chambers, and is significant above 35 °C (though not leading to efficiency loss). The gain correlates linearly with current, as shown on the right panel of Fig. \ref{fig_TPIcorr}, which matches with the output impedance with the high voltage unit. Since all these correlations (T, P, I) could be quantified, the measured gain has been corrected for, as shown on the top panel of Fig. \ref{fig_Lowflow_AllCH+TP+IH}. 

During the 40 days measurement at reduced flow, the gain was gradually decreasing and saturated at around 87\% of the original gain, which is a minor reduction leading to no visible tracking or trigger efficiency loss (as concluded in Section \ref{sec2.2}). It is also worth mentioning that the last chamber in the gas line, which received gas from the buffer every day due to breathing, was not contaminated significantly compared to the other chambers (Fig. \ref{fig_Lowflow_AllCH+TP+IH} top, blue data).

\begin{figure}
\includegraphics[width=9cm]{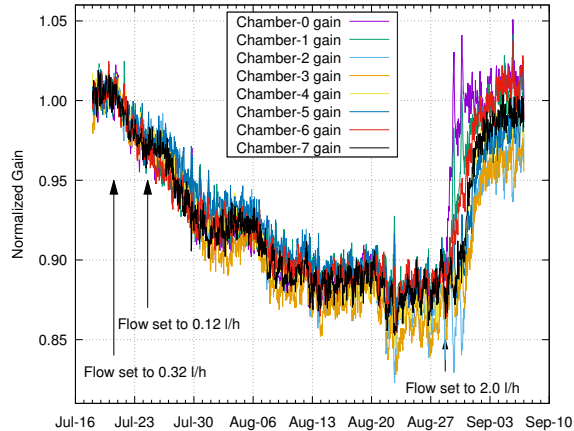}
\caption{\label{fig_Lowflow_CHbyCH} Normalized gains (relative to the mean of the first 5 days) for the individual chambers. High gas flow (2 l/h) was restarted on 29th Aug, after which gas cleaning proceeds in all chambers.
}
\end{figure}

The evolution of the corrected normalized gain chamber-by-chamber (Fig. \ref{fig_Lowflow_CHbyCH}) corresponds to the expectation from the lab measurements, with a gradual decreasing trend due to outgassing. At this flow value, the equilibration time is long: only 3 liters of gas are supplied each day, or in other words, it takes 1 week for the full gas exchange of a single chamber. Fig. \ref{fig_Lowflow_AllCH+TP+IH} top panel shows the mean of all detector chambers; if however one looks at the individual chambers in Fig. \ref{fig_Lowflow_CHbyCH}, it is apparent that all chamber gains move together. This rules out a uniform outgassing pattern (chambers later in the chain should have lower gains), an effect which we identified (after the measurements) to be due to the outgassing of the long input supply line. The gas line integrity is demonstrated by the sequential cleaning of the chamber after switching to the standard 2 l/h flow on 29th Aug. 

The measurements indicate that this detector system can be operated safely and without performance loss at a flow of 0.12 l/h (3 l/day) in ambient field conditions with the described gas buffering scheme.

\section{\label{sec6}Implementation examples}

\begin{figure}
\includegraphics[width=9cm]{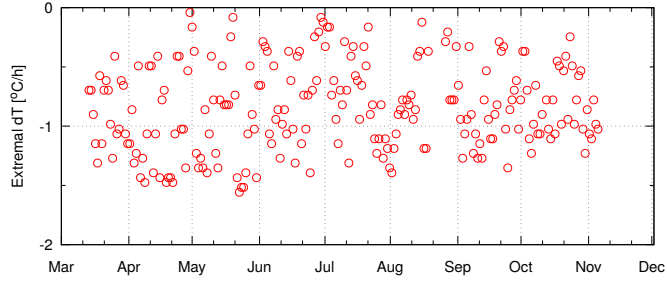}
\caption{\label{SMO_external_dT} SMO daily extremal negative temperature derivative over nearly a year. }
\end{figure}

\begin{figure}
\includegraphics[width=9cm]{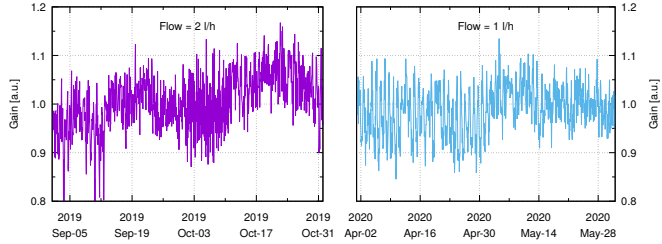}
\caption{\label{SMO_gain-time} SMO MMOS-09 gain stability: left panel shows running at 2 l/h, right panel running at 1 l/h (both over a 2 months period). }
\end{figure}

A practical implementation of the proposed structure is relevant at the Sakurajima Muography Observatory (SMO)\cite{SMO_2018, VCI_2020}. Currently there are 11 MWPC-based Muography Observation System (MMOS) modules, each containing 7 or 8 layers of 80 cm x 80 cm or 120 cm x 80 cm size chambers with a total sensitive area of 8.7 m$^2$. The maximum gas volume of an MMOS module is 160 l. The gas flow for one MMOS was 2 l/h until March 2020, after confirming safe operation the flow was reduced to 1 l/h per MMOS thus doubling the time between two gas cylinder replacement. Fig. \ref{SMO_external_dT} shows the highest daily negative temperature time derivatives at the SMO during a full year period. The drop rate reaches only -1.6 °C/h, that is, the systems will not ingest from the buffer volume even at 1 l/h, based on the measurement from Fig. \ref{fig_Phase}. However, for safety, buffer tubes were attached already to the end of every MMOS. The gain stability has been checked before and after March, as shown in Fig. \ref{SMO_gain-time}. 

It is important to note that at low flow, the detector outgassing is highly relevant, and each MMOS has to be properly checked for leaks or outgassing (see Section \ref{sec2.2}), a time consuming activity which could not be performed in 2020. 

In case of underground applications, the daily temperature variations are usually much smaller than on the surface, therefore the necessary buffer volume is smaller as well. In this case, the pressure change may need to be considered also. The axial diffusion can still be compensated with 10 m extra constant diameter buffer assuming non-zero input flow, and Eq. (\ref{eq_buffer}) can be used to calculate buffer volume, with the relative temperatures exchanged to relative pressures:

\begin{equation}\label{eq_pressure}
    V_{buffer} = V_{0} \cdot \left(P'_{min} - P'_{max} \right).
\end{equation}

For example, the case of 20 mbar pressure rise in less than 24 hours (e.g., typical central European strong cold front) means a 2\% relative pressure change and thus around 3 liters of required buffer for a 160 l volume detector with very low flow.

\section{Conclusions}

The present paper discusses one of the most relevant challenges of gaseous tracking detectors for muography: reducing gas consumption. In a practical implementation, the daily temperature variation causes a "breathing" effect, which is difficult to mitigate, and none of the trivial possibilities are attractive to designers (such as reduction of the detector volume, or highly effective heat insulation, or allowing considerable pressure change inside the detector chambers). We have investigated the application of a sufficiently long buffer tube, which is a simple and robust solution provided that some design specifications (length, volume, outgassing) discussed in the paper are followed. The concept was demonstrated during a long term (five weeks) outdoor measurement at low (3 l/day) flow with the same type of detectors as installed at the Sakurajima Muography Observatory.

\begin{acknowledgments}
This study was supported by the Joint Usage Research Project (JURP) of the University of Tokyo, Earthquake Research Institute (ERI) under project ID 2020-H-05, the "INTENSE" H2020 MSCA RISE project under Grant Agreement No. 822185, and the Hungarian NKFIH research grant under ID OTKA-FK-135349. The authors are grateful to Mr. \'Ad\'am Gera for his valuable support on engineering and fluid dynamics simulations.
\end{acknowledgments}

\section*{DATA AVAILABILITY}
The data that support the findings of this study are available from the corresponding author upon reasonable request.

\nocite{*}

\bibliography{aipsamp}

\end{document}